\documentclass[a4paper,10pt]{aastex62}

\hypersetup{linkcolor=red,citecolor=blue,filecolor=cyan,urlcolor=magenta}
\usepackage{natbib}
\usepackage{amsmath}
\usepackage{amsthm}
\usepackage{amssymb}
\usepackage{amsmath}
\usepackage{bm}
\usepackage[utf8]{inputenc}
\usepackage{graphicx}
\usepackage[caption=false]{subfig}
\usepackage{soul}
\usepackage{dblfloatfix}    

\usepackage{soul}
\mathchardef\mhyphen="2D

\shorttitle{Expanding Box Model PiC simulation of Heat Flux Instabilities in the inner heliosphere}
\shortauthors{Micera et al.}

\begin{document}

\title{On the role of solar wind expansion as a source of whistler waves: scattering of suprathermal electrons and heat flux regulation in the inner heliosphere.}

\correspondingauthor{A. Micera}
\email{alfredo.micera@oma.be}

\author[0000-0001-9293-174X]{A. Micera}
\affil{Solar-Terrestrial Centre of Excellence - SIDC, Royal Observatory of Belgium, Brussels, Belgium.}
\affil{Centre for Mathematical Plasma Astrophysics, KU Leuven, Leuven, Belgium.}

\author[0000-0002-2542-9810]{A. N. Zhukov}
\affiliation{Solar-Terrestrial Centre of Excellence - SIDC, Royal Observatory of Belgium, Brussels, Belgium.}
\affiliation{Skobeltsyn Institute of Nuclear Physics, Moscow State University, Moscow, Russia.}

\author[0000-0003-3223-1498]{R.~A. L\'{o}pez}
\affiliation{Departamento de F\'{i}sica, Universidad de Santiago de Chile, Santiago, Chile}

\author[0000-0003-1970-6794]{E. Boella}
\affiliation{Physics Department, Lancaster University, Lancaster, UK.}
\affiliation{Cockcroft Institute, Daresbury Laboratory, Warrington, UK.}

\author[0000-0003-3223-1498]{A. Tenerani}
\affiliation{Department of Physics, The University of Texas at Austin, Austin, USA.}

\author[0000-0003-3223-1498]{M. Velli}
\affiliation{University of California Los Angeles, Department of Earth, Planetary, and Space Sciences, Los Angeles, USA.}

\author[0000-0002-3123-4024]{G. Lapenta}
\affiliation{Centre for Mathematical Plasma Astrophysics, KU Leuven, Leuven, Belgium.}

\author[0000-0002-5782-0013]{M. E. Innocenti}
\affiliation{Institut f\"{u}r Theoretische Physik, Ruhr-Universität Bochum, Bochum, Germany.}

\begin{abstract}
\noindent The role of solar wind expansion in generating whistler waves is investigated using the EB-iPic3D code, which models solar wind expansion self-consistently within a fully kinetic semi-implicit approach. The simulation is initialized with an electron velocity distribution function modeled after Parker Solar Probe observations during its first perihelion at 0.166 au, consisting of a dense core and an anti-sunward strahl. This distribution function is initially stable with respect to kinetic instabilities. Expansion drives the solar wind into successive regimes where whistler heat flux instabilities are triggered. These instabilities produce sunward whistler waves initially characterized by predominantly oblique propagation with respect to the interplanetary magnetic field. The excited waves interact with the electrons via resonant scattering processes.
As a consequence, the strahl pitch angle distribution broadens and its drift velocity reduces. Strahl electrons are scattered in the direction perpendicular to the magnetic field, and an electron halo is formed. 
At a later stage, resonant electron firehose instability is triggered and further affects the electron temperature anisotropy as the solar wind expands.
Wave-particle interaction processes are accompanied by a substantial reduction of the solar wind heat flux. 
The simulated whistler waves are in qualitative agreement with observations in terms of wave frequencies, amplitudes and propagation angles.
Our work proposes an explanation for the observations of oblique and parallel whistler waves in the solar wind. We conclude that solar wind expansion has to be factored in when trying to explain kinetic processes at different heliocentric distances.

\end{abstract}


\section{Introduction}\label{sec.0}

Collisionless effects play a major role in the energy balance of the solar wind plasma \citep[e.g.][]{Marsch2006}. This is clear from the complex non-thermal shapes of the solar wind particle Velocity Distribution Functions (VDFs). The measurement of VDFs close to the Sun is crucial for two reasons.
First, VDFs may carry signatures of the mechanisms responsible for the solar wind heating and acceleration at the early stages of its evolution \citep{Ko1996, Bercic2020}. Second, information on the VDFs may be key to understanding the processes affecting the further evolution of the solar wind plasma during its propagation to 1 au and beyond. Observations by the Parker Solar Probe (PSP) mission \citep{Foxetal2016}, that are for the first time carried out below 0.3 au, contribute to our understanding of these processes. In particular, the SWEAP instrument (Solar Wind Electrons Alphas and Protons, see \citet{Kasper2016}), provides a unique view of electrons in the near-Sun environment. Electrons are the main contributors to the heat conduction that is one of the processes regulating the flow of energy in the solar wind \citep[e.g.][]{Scime1994}.

During its first perihelion PSP was immersed in a slow but highly Alfvénic solar wind stream emerging from a small equatorial coronal hole \citep{Bale2019}. There, it detected electron VDFs composed of a cold dense core, and a tenuous and hotter population, called ``strahl" (German for ``beam"), which streams outward from the Sun along the interplanetary magnetic field \citep{Halekas2020, Bercic2020}. In the solar wind another suprathermal population of electrons is usually observed: the halo, which is strongly non-Maxwellian, has higher temperatures and lower densities than the core, and it is distributed at all pitch-angles \citep{Verscharen2019_book}. 

Many processes may contribute to shape this peculiar, three component VDF. Since electrons are lighter than ions, solar gravity constrains them less effectively. High energy electrons are then able to escape from the Sun into interplanetary space. The electric field that arises extracts ions from the solar atmosphere as well \citep{Pierrard_1996,Maksimovic1997}. In addition, electrons are strongly shaped by the magnetic forces acting in the heliosphere. The divergence of the solar magnetic field and the simultaneous conservation of the adiabatic invariants have a focusing effect on the electrons and produce the strahl populations. On top of that, the ubiquitous influence of waves and collisions on solar wind particles shapes the electron VDF further.

Due to these marked asymmetries and non-thermal features, electrons carry the heat flux in the solar wind \citep{Feldman1975, Pilipp1987, Scime1994}. The electron heat flux bears information about the origin of the solar wind and the mechanisms that generate it in the solar corona. By observing the heat flux it is also possible to understand the topology of the interplanetary magnetic field. During Encounter 1, PSP detected a predominantly uni-polar anti-sunward heat flux \citep{Halekas2020b}, as one would expect to observe in the presence of a strahl going outward from the Sun along open magnetic field lines.
The heat flux also brings with it traces of various processes underlying particle-particle and wave-particle iterations that can significantly alter the structure of the electron VDFs. Indeed, by analyzing the heat flux in the solar wind at different heliocentric distances and in different solar wind regimes, it is possible to have an insight into the mechanisms that regulate it by reducing the skewness of the electron VDFs \citep{Scime1994}.

At 0.166 au, during PSP's first perihelion, the heat flux, almost entirely carried by the strahl, shows a clear anti-correlation with respect to the electron plasma $\beta$, the ratio of the electron thermal pressure to the magnetic pressure \citep{Halekas2020b}. No significant correlation is observed when one compares the heat flux to the collisonal age. This suggests that non-collisional mechanisms are playing a predominant role in regulating the electron heat flux during Encounter 1, rather than Coulomb collisions. 

At small heliocentric distances of PSP's first orbit, the halo fractional density is substantially lower than that of the strahl \citep{Halekas2020}, and lower than the already limited halo fractional density observed at 0.3 au by the Helios spacecraft \citep{Bercic2019}. This is consistent with observations made at different heliocentric distances, which show that while the halo fractional density increases, the strahl fractional density decreases when one moves further away from the Sun \citep{Maksimovic2005, Stverak_2009, Gurgiolo2012}. This also agrees with recent numerical models which reproduce the halo formation from the scattering of the strahl \citep{Vocks2005,Roberg-Clark2019, Tang2020, Micera2020b,Jeong-Seong2020}.

The data taken by the FIELDS experiment \citep{Bale2016} on-board PSP show that the inner heliosphere is permeated with a wide range of whistler fluctuations. \citet{Agapitov2020} and \citet{Cattell_2021_ArXive} found an appreciable presence of sunward whistler waves inside $0.3$ au, with intermixed propagation angles with respect to the magnetic field. Both the parallel and obliquely propagating whistler waves were characterized by large amplitudes. In addition, according to \citet{Cattell2021}, in the time intervals when whistler waves were observed, enhanced heat flux suppression rates were found, suggesting that the scattering of electrons by these waves may be a plausible mechanism to explain the regulation of the heat flux in the near-Sun solar wind.
\citet{Jagarlamudi2021} and \citet{Cattell2021} reported a significant broadening of the strahl pitch angle distribution in coincidence with the detection of short-duration whistler wave trains by PSP, confirming the fundamental role of the interaction between electrons and these waves in regulating the near-Sun solar wind heat flux. It is clear that understanding the origin of these waves and the consequences of their interactions with electrons may shed some light on a key mechanism responsible for the energy redistribution in the solar wind.
At 1 au, in addition to the oblique large-amplitudes whistler waves \citep{Breneman_2010, Cattell2020}, parallel small-amplitude whistler waves are also often observed \citep{Lacombe2014, Stansby2016, Tong2019}.

There may be various potential sources of propagating whistler waves. 
Whistlers can be driven by wave-wave interactions which lead to turbulent cascades and to power law spectra of the magnetic field in the whistler regime \citep{Stawicki2001,Smith2006, Tang2020}. However, observations of whistler fluctuations are found to be regularly concentrated in regions of the inner heliosphere where electrons are characterized by high electron $\beta$ values, or at least higher than those occurring in the close proximity \citep{Jagarlamudi2020, Cattell2021}. A positive correlation is also observed when one compares the values of electron $\beta$ with the magnitude of the pitch-angle scattering of the strahl \citep{Pilipp1987,Crooker2003, Bercic2019}, suggesting that kinetic instabilities self-generated by electrons are the possible source of the observed whistler waves and possible candidates for shaping the non-thermal features of the electron VDF.
Whistler waves can be also generated by the so-called whistler temperature anisotropy instability \citep{Gary1993}, which in turn is produced by a temperature anisotropy of the electrons with $T_{e, \perp} > T_{e, \parallel}$, where $\perp$ and $\parallel$ denote directions perpendicular and parallel to the magnetic field. However, this electromagnetic instability has the maximum growth rate parallel to the magnetic field, which would not explain the generation of the observed highly oblique whistler waves. Moreover, it turned out to have a negligible impact on the scattering of the high-energy components of the electron VDF\citep{Saito2007}. 
Whistler heat flux instabilities are the most promising mechanism to generate the whistler waves observed in the inner heliosphere.
Indeed, \citet{Cattell2020} found a clear dependence of the occurrence of the whistler waves on the threshold of whistler heat flux instabilities.
\citet{Halekas2020b} showed that during the first two PSP orbits the thresholds of whistler heat flux instabilities \citep{Vasko2019,Verscharen_2019} were fully bounding the observed electron heat flux values, and the strahl fractional densities were constrained below the limit values predicted by these instabilities. 

In a weakly collisional magnetized plasma, heat flux instabilities can be self-generated by counter-streaming populations of electrons. These conditions occur frequently in the solar wind, where the strahl presents often an anti-sunward drift with respect to the protons, and to maintain a balance in current, the core also drifts, with a much lower speed, in the direction opposite to that of the strahl \citep{Feldman1975,Scime1994}. Core and strahl, with their relative drift velocities, can self-induce microinstabilities which can reduce the heat flux below the limit values provided by collisional models \citep{Spitzer_1953}.
The nature of the heat flux instabilities can differ according to the plasma conditions. Depending on the drift velocity of the strahl, electromagnetic parallel \citep{Gary1975,Gary1977, Gary1994,Gary1999, Kuzichev2019,Lopez2019} or oblique \citep{Komarov2018,Horaites2018,Vasko2019,Verscharen_2019,Lopez2020, Micera2020b} whistler heat flux instabilities (WHFIs) can be excited. The free energy of the counter-streaming electron populations is converted into magnetic energy in the form of parallel right-hand circularly polarized or oblique right-hand elliptically polarized whistler-mode waves. A large variety of electrostatic or hybrid instabilities  \citep[e.g.][]{Gary1978,Shevchenko2010,Pavan2013, Shaaban2018MN, Lopez2020} can also be self-generated by the skewness of the electron VDF. However, these instabilities were found to have significantly lower growth rates than WHFIs and do not satisfy the resonant conditions to scatter the strahl into the halo \citep{Verscharen_2019}.
We therefore focus here on WHFIs, especially on the oblique WHFI, as a crucial source of whistler waves in the inner heliosphere.

The occurrence of microinstabilities in the solar wind is strictly related to the evolution of its bulk parameters that vary as it propagates in the heliosphere. If the solar wind expands spherically, then, in the absence of wave-particle interactions, the temperatures should follow the double adiabatic Chew–Goldberger–Low (CGL) model \citep{Chew1956}. According to this model, due to the conservation of the two adiabatic invariants $T_{\perp}/B$ and
$T_{\parallel}B^2/n^2$, in a spherically expanding solar wind, where both the radial magnetic field $B$ and the density $n$ decrease as $R^{-2}$, with $R$ being the heliocentric distance, a marked temperature anisotropy would result even starting from an initially isotropic distribution ($T_{\perp} \sim R^{-2}$ and $T_{\parallel} = const$).

However, due to kinetic instabilities, i.e. electron and proton firehose instabilities, and to the resulting wave-particle interactions that produce a simultaneous parallel cooling and perpendicular heating, these temperature trends are not observed \citep{Hellinger2008,Matteini2012, Innocenti2019, Micera2020}.
In this respect, \citet{Innocenti2020} highlighted the indirect role of solar wind expansion in heat flux regulation: solar wind expansion can trigger or modify the evolution of firehose instabilities, which can in turn contribute to heat flux regulation.

In the presence of multiple-component electron VDFs, firehose instabilities (triggered as function of temperature anisotropy and plasma $\beta$ values) are not the only instabilities that can alter the solar wind expansion trend: heat flux instabilities, triggered by the presence of counter-streaming electron populations, become a recurring possibility. Which specific heat flux instability is triggered depends on the properties of counter-streaming electrons.

For the whistler heat flux instabilities, the onset strongly depends on the ratio between the strahl drift velocity and the Alfvén velocity, $u_s /V_{A}$. When this ratio increases, the instability threshold becomes lower and its growth rate is boosted \citep{Verscharen_2019, Lopez2020}. In the inner heliosphere, where the magnetic field is predominately radial ($B \sim R^{-2}$), the Alfvén velocity decreases as $R^{-1}$. If the strahl drift velocity decreases with heliocentric distance slower than that, we may expect that the threshold of the instability also decreases during the solar wind passage through the inner heliosphere. This would mean that, as the solar wind propagates, it would encounter more favorable conditions for the quasi-continuous generation of unstable whistler-mode waves, as conjectured by \citet{Verscharen_2019}.

In this work, we study the role of the solar wind expansion in triggering electron kinetic instabilities and, hence, its role in regulating the heat flux evolution in the solar wind. We investigate if the solar wind expansion can drive the onset of whistler heat flux instabilities and hence contribute to the generation of whistler waves in the inner heliosphere. We perform a fully kinetic, Expanding Box Model Particle-in-Cell (PiC) simulation \citep{Innocentietal2019} and we infer how the impact of kinetic instabilities varies during the solar wind propagation and how efficient they are in limiting the heat flux at different heliocentric distances.

This paper is organised as follows. Section \ref{sec.2} describes the simulation framework and setup, presents an overview of the code, the parameters used, and initial conditions. In Section \ref{sec.3} we present and explain the main results. In Section \ref{sec.4} we discuss the implications of our results for the understanding of kinetic instabilities in the solar wind and for the interpretation of in situ observations.
Section \ref{sec.5} summarizes our results and reports the conclusions.

\section{Numerical Model}\label{sec.2}
We study the onset and evolution of microinstabilities in the expanding solar wind plasma self-consistently with EB-iPic3D simulations. The EB-iPic3D code \citep{Innocentietal2019} introduces for the first time the Expanding Box Model (EBM) \citep{Velli1992,Grappin1996,Tenerani2017} into a semi-implicit fully kinetic PiC framework \citep{Brackbill1982,Lapenta2006,Markidis2010, INNOCENTI20173}. This allows studies of the interplay between the solar wind plasma expansion and kinetic processes down to the electron scales. In the EBM, a coordinate transformation is used to incorporate expansion effects in the evolution equations for fields and particles. This gives the possibility of reducing the size of the computational domain and hence model the solar wind expansion at an affordable computational cost \citep{Innocentietal2019,Innocenti2019}. In addition, in semi-implicit PiC codes, the strict constraints on the temporal and spatial scales needed to keep explicit PiC codes stable are almost completely removed \citep{Lapentaetal2017, Gonzalez-Herrero2018, Micera2020}. These two innovative aspects together allow studies of the solar wind dynamics from a kinetic point of view and to explore the long-term evolution and large-scale effects of the expansion. 

In the EBM, one follows a radially expanding solar wind plasma parcel as it moves away from the Sun in the comoving frame.
The average heliocentric distance of the box center is therefore given by $R(t) = R_0 + u_0~t$, where $R_0$ is the initial distance from the Sun and $u_0$ is the constant radial velocity at which the solar wind moves away from the Sun. The value $\tau_{exp} = R_0 / u_0$ defines the time scale of the expansion and thus governs the rate at which the simulation parameters vary with the heliocentric distance. All the expansion related quantities can be expressed via a parametric dependence on $R(t)$.

We investigate the interaction between the adiabatic expansion and the development of kinetic instabilities. Our objective is to follow how expansion can lead to the generation of electromagnetic fluctuations that affect particle VDFs. To do so, we initialise our simulation with realistic plasma conditions, observed by PSP at its first perihelion, and we follow the non-linear evolution of the electron VDF during its interaction with the waves that are generated by the electrons themselves as the solar wind expands. 

We carry out a two spatial dimensions, three velocity components (2D3V) fully kinetic EBM simulation. The electron distribution function initially consists of two populations. The strahl (subscript ``s") has a density $n_s$ that is only $5 \%$ of the total electron density $n_e = 350$ cm$^{-3}$ and a drift velocity $u_s$ that is directed anti-sunward and, in agreement with observations, equals $6900$ km/s \citep{Halekas2020}. The core (subscript ``c") contains the remaining $95\%$ of the electron density and has a sunward drift $u_c$ relative to the proton rest frame, in order to satisfy zero-current condition ($n_{c}~ u_{c} + n_{s}~ u_{s}~ = 0$) and hence to ensure the quasi-neutrality of the plasma.
In agreement with recent PSP observations \citep[i.e][]{Halekas2020,Bercic2020}, we adopted an initially isotropic core with $k_B T_c = 30$ eV, where $k_{B}$ is the Boltzmann constant. The strahl is characterised by a parallel temperature, $k_B T_{s \parallel} = 173$ eV, twice larger than the perpendicular temperature, in order to take into account its limited angular extent. The ions, assumed to be only protons, have the real proton-to-electron mass ratio $\mu =~m_i / m_e =~1836$, a zero drift velocity in their reference frame, density $n_i=n_e$, and are initially isotropic with $T_i = T_c$. The core and strahl initial VDFs are described via drifting-Maxwellian and drifting-bi-Maxwellian distributions, respectively:

\begin{equation}
     \begin{aligned}
f_j (v_{\parallel}, v_{\perp}, t=0) = \frac{(2 \pi)^{- 3/2}}{w_{j \perp}^2 w_{j \parallel }} \exp \left(-\frac{v_{\perp}^2}{2 w_{j \perp}^2} - \frac{(v_{\parallel} - u_j)^2}{2 w_{j \parallel}^2} \right), 
    \label{sss}   
     \end{aligned}     
\end{equation}
with $w_{j} = \sqrt{k_B T_j / m_j}$ and $u_j$ being respectively the thermal and drift velocities of the species $j$ (core electrons, strahl electrons and protons). 

The initial background magnetic field is uniform and directed anti-sunward, $\bm{B}_0=B_0 \hat{e}_x$, with $B_0 = 60$ nT, so that the initial proton Alfvén speed $v_{A} = B_0/ \sqrt{4 \pi n_e m_i} = 0.00023\, c$, with $c$ being the speed of the light in vacuum ($x$ is the parallel spatial coordinate, and $y$ the perpendicular one). The boundary conditions adopted are periodic for both particles and fields.

The simulation domain sizes are $L_x = L_y = 8\, d_i$, resolved with $800 \times 800$ cells, where $d_i = c / \omega_{pi}$ is the ion skin depth, and $\omega_{pi} = \sqrt{4 \pi e^2 n_i / m_i}= 24615$ s$^{-1}$ the ion plasma frequency (at the beginning of the simulation), with $e$ being the elementary charge. We use $1024$ particles per cell per species (the core and strahl electron populations are treated as separated species in the code). The temporal step is $\Delta t = 0.05\, \omega_{pi}^{-1}$, and the simulation runs up to the time $t = 16500\, \omega_{pi}^{-1}$. 
With these parameters the initial relevant frequency ratios are: $\omega_{pe} / \omega_{ce} = 100$, $\omega_{pi} / \omega_{ci} = 4284.85$ and $\omega_{pi} / \omega_{ce} = 2.33$, with $\omega_{pe} = \sqrt{4 \pi e^2 n_e / m_e}$ being the electron plasma frequency and $\omega_{cj} = e B_0 / c\, m_j$ is the gyrofrequency of the species $j$. For the spatial scales one should consider that: $d_i / \rho_{e} = 56$, and $d_i / d_{e} = 42.85$  with $\rho_e = w_e / \omega_{ce}$ being the electron gyroradius and $d_e = c / \omega_{pe}$ is the electron skin depth.

An important choice in our simulation is the expansion time $\tau_{exp}$, and its ratio with respect to the e-folding time ($\tau$) of the instabilities we aim to study.
On the one hand, this ratio must be large enough to keep these two timescales well separated. On the other hand, the effects of the expansion must be fast enough to make the simulation feasible within reasonable computation resources. 
The selection of $\tau_{exp}$ was done on the basis of the typical growth rates of whistler heat flux instabilities for characteristic solar wind parameters. To get an estimate of this value, we relied on our previous work \citep{Micera2020b}, and on the results obtained by \citet{Lopez2020} who solved the linear dispersion relation for a large range of plasma parameters. This was then followed by a convergence study where simulations performed with different resolutions and expansion times confirmed the results of the present study. We thus chose $\tau_{exp} = 10000\, \omega_{pi}^{-1}$, which gives $\tau_{exp} / \tau_{~OWHFI}\approx 30$, where $\tau_{~OWHFI} = 1/\gamma_{~OWHFI}$ is the inverse of the maximum growth rate $\gamma_{~OWHFI}$ of the oblique WHFI studied in \citet{Micera2020b}. $\tau_{exp}$ is lower than the realistic solar wind expansion time, but still large enough to keep the timescales of the instability and of the expansion well separated, hence our simulation results are physically significant. 
We obtained this expansion time by maintaining the radial solar wind velocity consistent with PSP observations during Encounter 1 ($u_0 = 540$ km/s = $0.0018\, c$), but assuming a very small initial heliocentric distance of the solar wind co-moving simulation box ($R_0 = 18~d_i$). It should be kept in mind that $u_0$ and $R_0$ are free parameters and only their ratio regulates how fast our plasma expands. This small $R_0$ is merely a convenient value that gives a characteristic expansion time that fits the criteria outlined.

The energy evolution in the simulation follows the expected adiabatic trend with heliocentric distance. For example, when the initial magnetic field is purely radial, the magnetic energy evolves as $E_B \sim B^{2} \sim R^{-4}$. Magnetic energy fluctuations constitute only a small fraction of the total magnetic energy (see Section \ref{sec.3.1}).

\section{Results}\label{sec.3} 

\subsection{PiC simulation of wave-particle resonant interactions}\label{sec.3.1} 

\begin{figure*}
 \centering
 \includegraphics[width=0.8\textwidth]{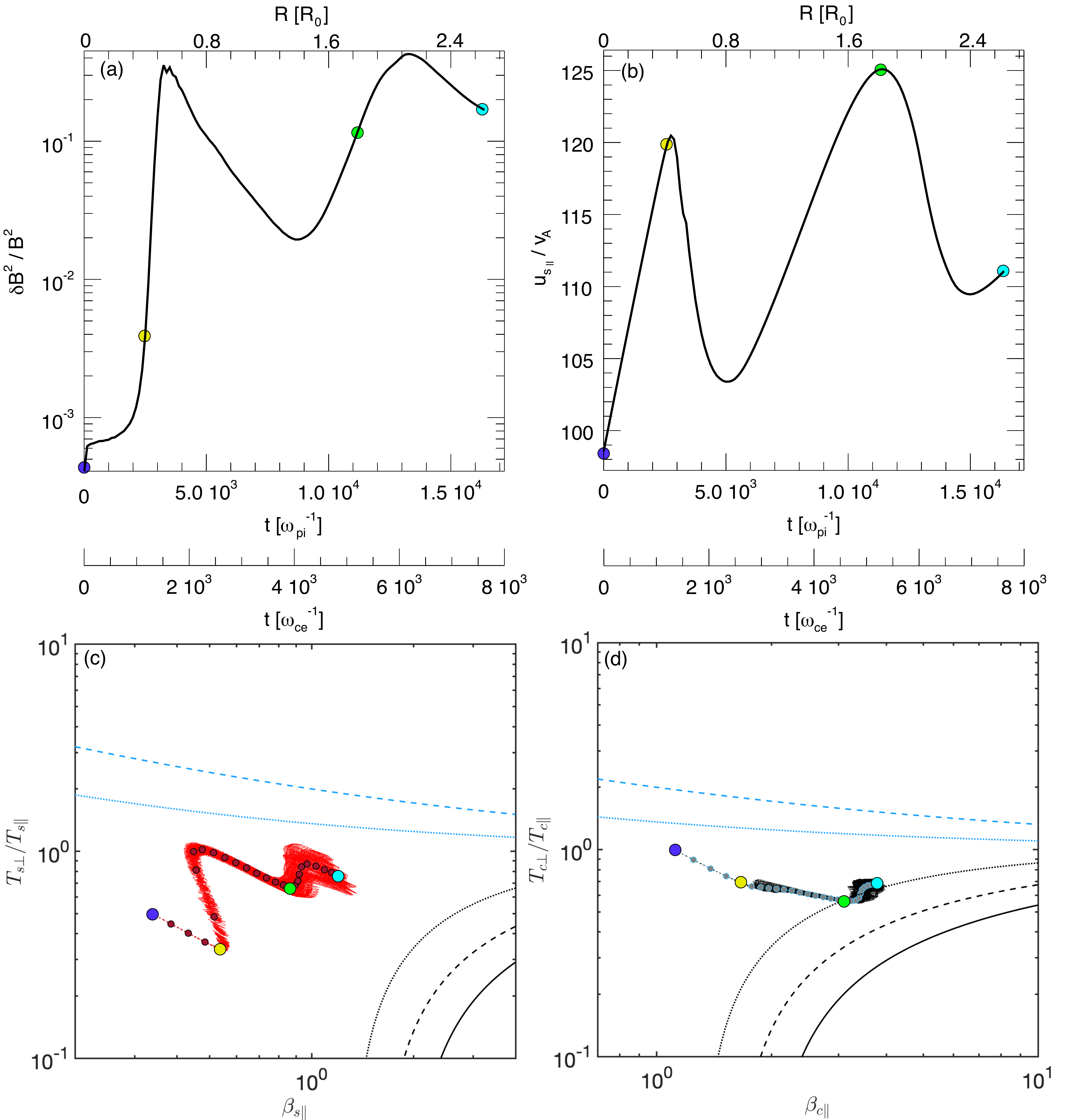}
 \caption{Total magnetic energy fluctuations normalized to the the mean magnetic field energy as a function of time (in the units of $\omega_{pi}^{-1}$ and $\omega_{ce}^{-1}$) and heliocentric distance $R/R_0$ (panel (a)).
 Parallel strahl drift velocity normalized to the local Alfvén velocity as a function of time and heliocentric distance (panel (b)).
 Strahl (panel (c)) and core (panel (d)) trajectories in the $\beta_{\parallel}$ vs $T_{\perp}/T_{\parallel}$ plane.
 Solid, dashed, and dotted black curves represent isocontours of the oblique EFI for $\gamma = 0.2, 0.1$ and $0.01~\omega_{ce}$, respectively \citep{Gary2003}.
 The dashed and dotted blue curves are the isocontours of growth rates $\gamma = 0.1$ and $0.01~\omega_{ce}$ for the whistler temperature anisotropy instability, respectively \citep{GaryandWang1996}. Red and gray filled dots in panels (c) and (d) correspond to temperature and $\beta$ values averaged over the whole simulation box every $625\, \omega_{pi}^{-1}$, while the background shadings depict the spread of the simulated data values.
 In all the four panels the initial time is marked with a blue dot, the first onset of the oblique WHFI ($t = 2500\, \omega_{pi}^{-1}$) with a yellow dot, the oblique WHFI + EFI second growth stage ($t = 11250\, \omega_{pi}^{-1}$) with a green dot, and the final simulation time with a cyan dot.} \label{fig.1}
\end{figure*}

We first verified, both through simulation and linear theory, that our plasma system with realistic initial conditions described in Section \ref{sec.2} is stable with respect to any kinetic instability if it does not expand.
For the expanding system, Figure~\ref{fig.1}(a) shows that the total magnetic fluctuation energy normalized to the mean magnetic field energy $\delta B^{2} / B^2$ peaks at $t \approx 3200\, \omega_{pi}^{-1}$ after a fast growth phase. Then it decreases significantly until secondary instabilities are triggered. A second peak is reached at $t \approx 13000\, \omega_{pi}^{-1}$, and finally the energy ratio relaxes again until the end of the simulation. Due to the expansion, the plasma parcel comoving with the solar wind becomes first unstable to the oblique WHFI \citep{Vasko2019,Verscharen_2019,Lopez2020,Micera2020b}, which is responsible for the first exponential rise of the magnetic energy fluctuations. After these fastest growing modes are saturated and the whistler instability completely relaxed, the energy of the magnetic field fluctuations monotonically decreases. It then grows again when new modes are destabilized due to the variation of plasma parameters caused by the further plasma expansion.

The adiabatic expansion of the plasma produces a decrease of the core-strahl differential velocity \citep{Innocenti2020}. This decrease is however less steep than that of the local Alfvén speed, adiabatically evolving as $R^{-1}$ in the undisturbed solar wind with predominantly radial interplanetary magnetic field. Consequently, the ratio $u_{s \parallel} / v_A$ as function of the time, displayed in Figure~\ref{fig.1}(b), increases in the first stage of the simulation, when the plasma expands following a double-adiabatic evolution, until it reaches the oblique WHFI threshold. The instability onset produces a further, sudden decrease of the strahl drift velocity and inverts the trend of the ratio evolution.
Thus, two concurrent mechanisms take place: on the one hand, continuous expansion and the resulting excitation of whistler-mode waves, and, on the other hand, significant regulation of the strahl drift velocity by the generated waves and the subsequent saturation of the instability. The oblique WHFI goes through cycles of stabilization and destabilization, with effects clearly reflected in the evolution of $u_{s \parallel} / v_A$.

In Figures~\ref{fig.1}(c), (d) we display, respectively, the tracks of the strahl and core electron populations in the $\beta_{j \parallel}$ vs $T_{j \perp}/T_{j \parallel}$ plane, with $\beta_{j \parallel} = 8 \pi n_j k_B T_{j \parallel} / B^2$. Figure~\ref{fig.1}(c) confirms that the strahl starts from a quasi-stable state, far from microinstability thresholds. After a first phase of adiabatic expansion, the strahl enters in the cycle that leads to the redistribution of the its kinetic energy from the parallel to the perpendicular direction and vice versa, depending on whether the oblique WHFI is in a growth or relaxation phase. The core also deviates from the double adiabatic evolution when the first WHFI occurs (Figure~\ref{fig.1}(d)). However, its largest contribution to the system evolution occurs in the last stages of the simulation, when its track crosses the threshold of the oblique electron firehose instability (EFI) \citep{LiandHabbal2000,Gary2003,Camporeale2008,Shaaban2019c,Innocenti2019} and is successively bounced back to the stable region. In this way, the core significantly participates in the generation of the late peak in the evolution of the fluctuating magnetic energy of Figure~\ref{fig.1}(a). Thereby, the late increase in energy can be traced back to an almost simultaneous stimulation of the oblique EFI by the core and a secondary oblique WHFI by the strahl (this is confirmed by the results of the linear dispersion analysis, see Section \ref{seclin-th}). We note that in both Figures~\ref{fig.1}(c) and (d) there is no spread around the average values of temperature anisotropy and $\beta_{\parallel}$ during the adiabatic phase. The oscillations around the mean values increase when instabilities develop, especially in the last stages of the simulation, when oblique WHFI and EFI are triggered together.

\begin{figure*}
 \centering
 \includegraphics[width=1\textwidth]{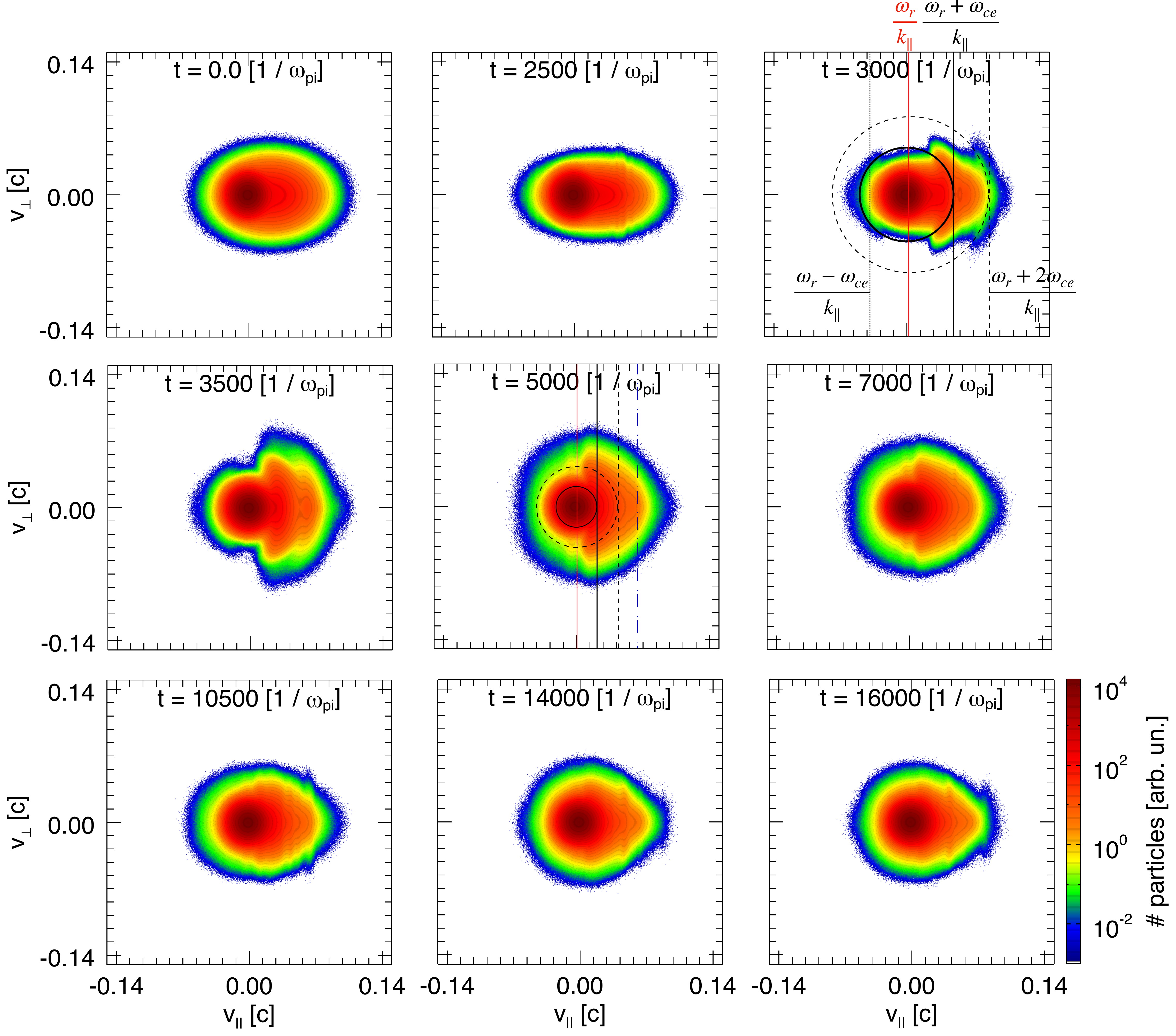}
 \caption{Total electron distribution functions in the $v_{\parallel} - v_{\perp}$ phase space at $t=0$ (panel (a)), $t = 2500$ (oblique WHFI first onset, panel (b)), $t = 3000$ (oblique WHFI linear growth phase, panel (c)), $t = 3500$ (oblique WHFI non-linear relaxation phase, panel (d)), $t = 5000$ (oblique WHFI non-linear relaxation phase, panel (e)), $t = 7000$ (WHFIs saturation, panel (f)), $t = 10500$ (EFI onset, panel (g)), $t = 14000$ (oblique WHFI + EFI non-linear relaxation phase, panel (h)), $t = 16000$ (final distribution, panel (i)). The time is in units of $\omega_{pi}^{-1}$.
 Black and blue vertical lines in panels (c) and (e) indicate $v_{\parallel}$ values at which gyroresonances of whistler waves with electrons are expected. They are plotted at $v_{\parallel} =-0.041  ~c$ in panel (c) and $v_{\parallel} =-0.02  ~c$ in panel (e) for $n=-1$, at $v_{\parallel} =0.0457  ~c$ in panel (c) and $v_{\parallel} =0.0218  ~c$ in panel (e) for $n=1$, at $v_{\parallel} =0.089  ~c$ in panel (c) and $v_{\parallel} =0.0429  ~c$ in panel (e) for $n=2$, at $v_{\parallel} =0.064  ~c$ in panel (e) for $n=3$.
 Red vertical lines, drawn at $v_{\parallel} =0.00233  ~c$ in panel (c) and $v_{\parallel} =0.0007  ~c$ in panel (e), respectively delineate the electron population that fulfills the $n=0$ Landau resonant condition with whistler waves.
 The circles show electrons diffusion paths due to the anomalous cyclotron resonance interaction as constant energy surfaces in the frame moving with the parallel phase velocity of the wave $v_{ph}= \omega_{r}/ k_{\parallel}$. They are centered at $v_{\parallel} = v_{ph}$, $v_{\perp} =0$, and their radius equals $n~ \omega_{ce}/ k_{\parallel}$.} \label{fig.2}
\end{figure*}

\begin{figure*}
 \centering
 \includegraphics[width=0.75\textwidth]{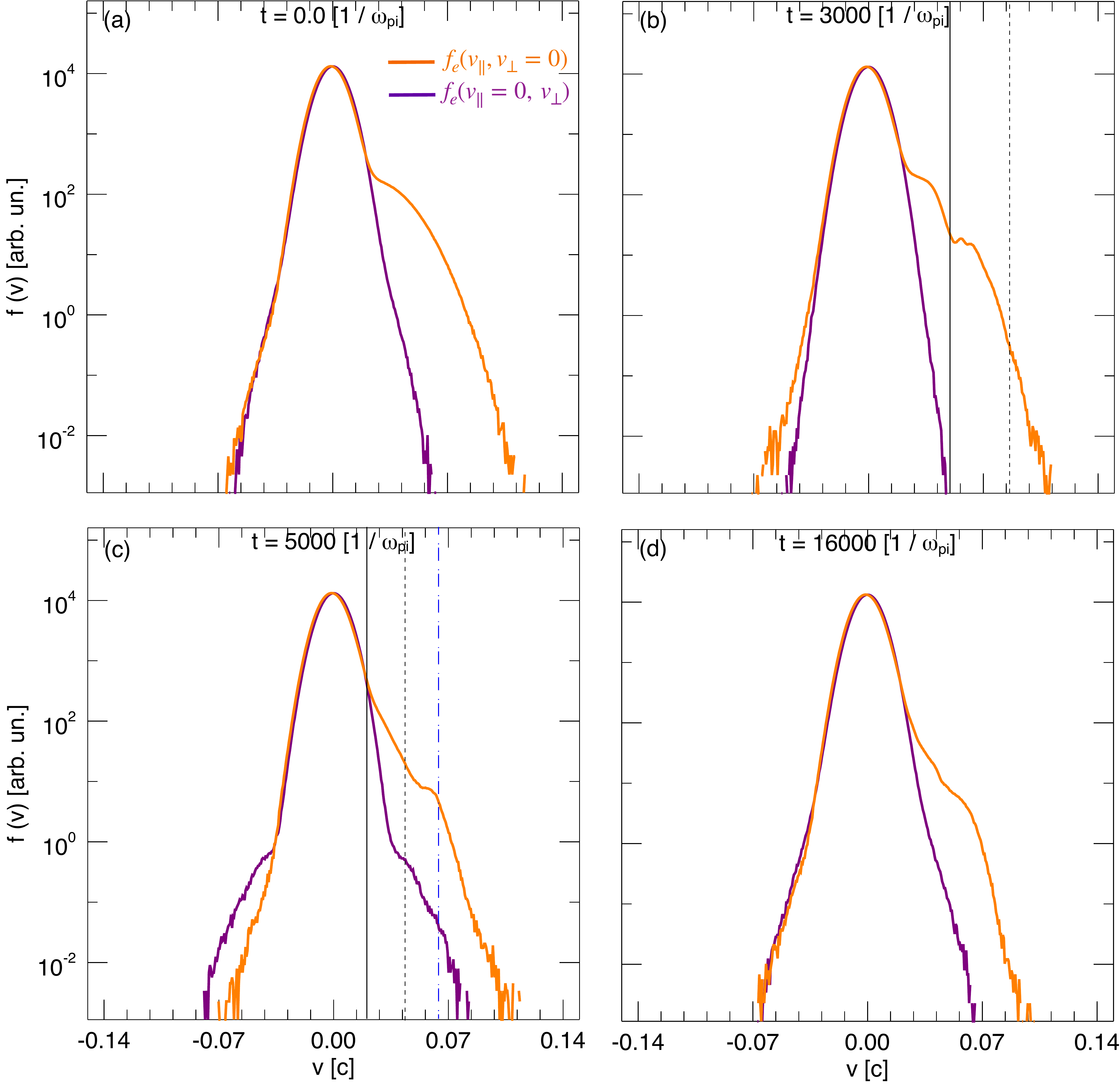}
 \caption{Total electron VDF cuts along the parallel ($f_e (v_{\parallel} , v_{\perp} =0$), orange solid line) and perpendicular ($f_e (v_{\parallel} =0, v_{\perp}$), purple solid line) directions at $t=0$ (panel (a)), $t = 3000\, \omega_{pi}^{-1}$ (panel (b)), $t = 5000\, \omega_{pi}^{-1}$ (panel (c)) and $t = 16000\, \omega_{pi}^{-1}$ (panel (d)).
 Vertical lines in panels (b) and (c) are drawn at the intersections of the resonant surfaces of $n=1, 2, 3$ with the $v_{\parallel} =0$ axis (see Figures~\ref{fig.2}(c) and (e)).}\label{fig.3}
\end{figure*}

In Figure~\ref{fig.2} we show the total electron VDF as a function of $v_{\parallel}$ and $v_{\perp}$ at important stages of its evolution.
The cuts of the total electron VDF $f_e (v_{\parallel}, v_{\perp} =0)$ and $f_e (v_{\parallel} =0, v_{\perp})$ are shown in Figure~\ref{fig.3}.
The excitation of microinstabilities has noticeable effects on the electron distribution function. The generated waves resonantly interact with the electron VDF, modifying its bulk parameters as well as its shape. We depict in Figure~\ref{fig.2}(a) the initial, stable electron VDF. Between $t = 0$ and $t = 2500\,\omega_{pi}^{-1}$ adiabatic expansions reduces the perpendicular velocity of the distribution. For the strahl, this can be interpreted as adiabatic focusing in the expanding solar wind.  

At $t = 2500\, \omega_{pi}^{-1}$, the VDF becomes unstable to WHFI, Figure~\ref{fig.2}(b). At this point the triggered whistler waves start to resonate with the electrons producing the ``horns" that are visible at $v_{\parallel} > 0$, which correspond to a deviation of the suprathermal electron population from its original bi-Maxwellian shape. The interaction of the electrons with the whistler waves, is accentuated at $t = 3000\, \omega_{pi}^{-1}$ (Figure~\ref{fig.2}(c)), almost at the peak of the linear growth phase of the oblique WHFI, producing multiple signatures of resonant interaction. A significant transfer of electron momentum from the parallel to the perpendicular direction and a considerable reduction of the strahl drift velocity can be observed, which lead to a simultaneous broadening of the strahl pitch angle distribution. Electrons with $v_{\parallel}$ satisfying the condition of resonance with the growing whistler waves are scattered towards higher values of $v_{\perp}$, forming a nascent halo that characterises the distribution function during the whole propagation of the solar wind through the heliosphere.

\begin{figure*}
 \centering
 \includegraphics[width=1\textwidth]{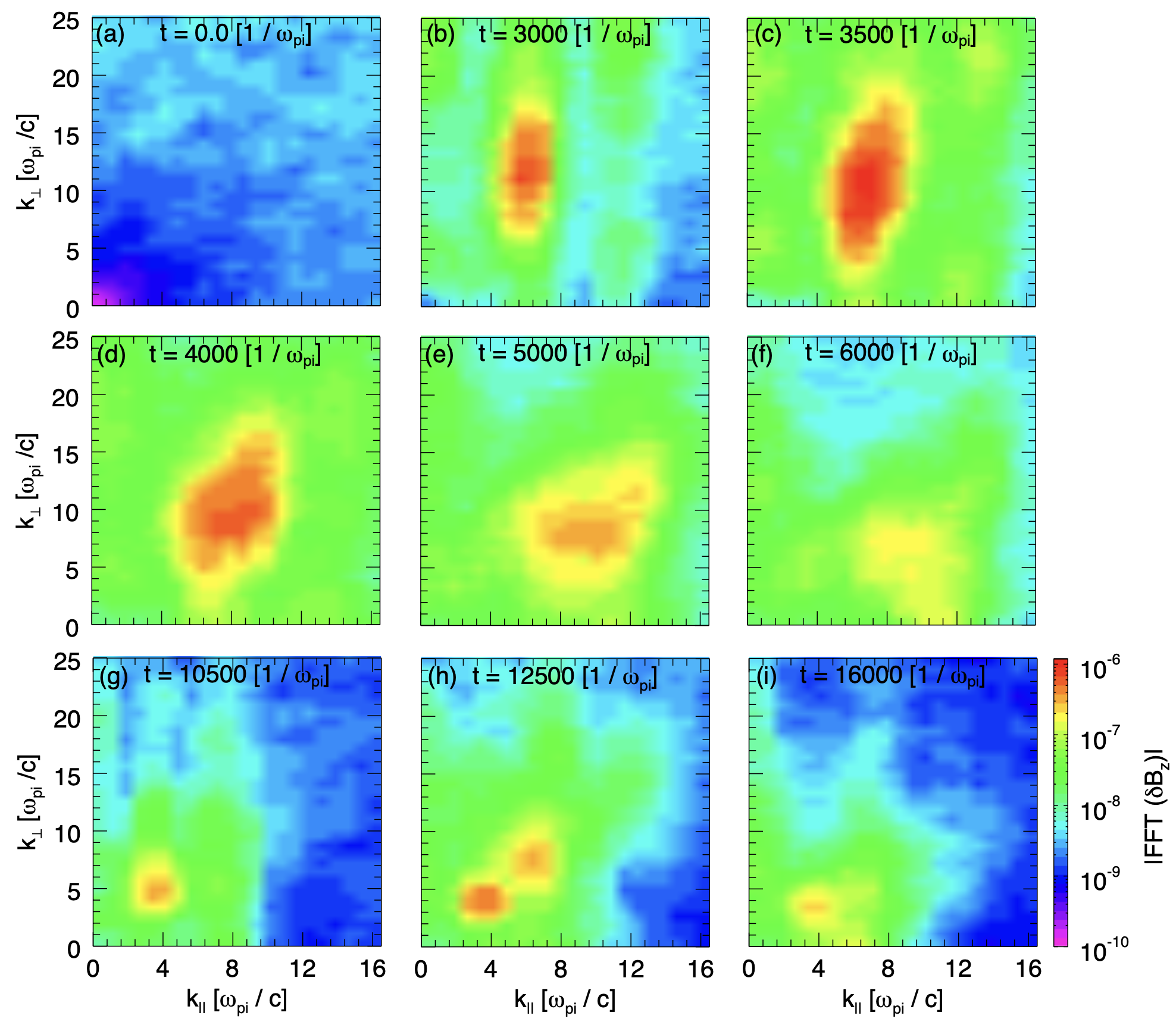}
 \caption{Fast Fourier transform of the out-of-plane magnetic field fluctuations ($\text{FFT}(\delta B_z)$) at $t=0$ (initial power spectrum, panel (a)), $t = 3000$ (oblique WHFI linear growth phase, panel (b)), $t = 3500$ (oblique WHFI non-linear relaxation phase, panel (c)), $t = 4000$ (oblique WHFI non-linear relaxation phase, panel (d)), $t = 5000$ (oblique WHFI non-linear relaxation phase, panel (e)), $t = 6000$ (WHFI saturated marginal state, panel (f)), $t = 10500$ (EFI onset, panel (g)), $t = 12500$ (oblique WHFI + EFI linear growth phase, panel (h)), $t = 16000$ (final power spectrum, panel (i)). The time is in units of $\omega_{pi}^{-1}$.\\} \label{fig.4}
\end{figure*}

In Figure~\ref{fig.2}(c) vertical lines mark the $v_{\parallel}$ values at which electrons are expected to interact via cyclotron, Landau and anomalous cyclotron resonances with the whistler waves which at that time permeate the simulated plasma.
The whistler resonance condition with electrons is given by $v_{\parallel} = (n~ \omega_{ce} + \omega_{r}) /k_{\parallel}$, with $\omega_r$ and $k_{\parallel}$ being respectively the real wave frequency and the parallel wave number of the fastest growing mode at a given time, with $n = -1$ for the cyclotron resonance, $n = 0$ for the Landau resonance, and $n\geq 1$ for the anomalous cyclotron resonances \citep{Krall1973}. The circles, centered at $v_{\parallel} = \omega_{r} /k_{\parallel}$ and $v_{\perp} =0$, illustrate the diffusion paths of the electrons that undergo anomalous cyclotron resonant interactions, as constant energy surfaces in the reference frame of the wave \citep{Roberg-Clark2019, Verscharen_2019}. We compute $\omega_{r}$ and $k_{\parallel}$, with their ratio giving the parallel phase velocity of the waves ($v_{ph} = \omega_{r} / k_{\parallel}= 0.00233~c$, with $\omega_{r}= 0.014~ \omega_{pi}$ and $k_{\parallel}=6~ \omega_{pi}/c$), via the linear dispersion relation (see Figure~\ref{fig.4b}(b) below) and fast Fourier transform (FFT) in space of the out-of-plane component of the fluctuating magnetic field $\delta B_z$ (see Figure~\ref{fig.4}(b) below), respectively. 

One can see that the $n=1$ and $n=2$ resonant surfaces intersect quite accurately the two large horn-like structures that appear in the frontal/anti-sunward region of the electron VDF during the evolution of the oblique WHFI (Figure~\ref{fig.2}(c)). This implies that a portion of the strahl electron population, characterized by a $v_{\parallel}$ that fulfills the anomalous cyclotron resonance conditions is scattered by whistler waves towards higher perpendicular velocities. This leads to the isotropization of the suprathermal electrons and to the formation of a nascent electron halo. Due to their elliptical polarization, the oblique whistlers modes can trigger cyclotron resonances with both positive and negative $n$ \citep{Komarov2018}. Thus, electrons in the $v_{\parallel} < 0$ region of the VDF are at this stage efficiently scattered by the $n=-1$ cyclotron resonance (Figure~\ref{fig.2}(c)).

Shortly after the end of the linear growth phase of the oblique WHFI, at $t = 3500\, \omega_{pi}^{-1}$, the halo formation is still restricted mainly to the portion of strahl with velocities satisfying the anomalous cyclotron resonant conditions. The nascent halo appears as a marked deformation of the distribution function, as depicted in Figure~\ref{fig.2}(d).
However, this does not last long as the electron VDF relaxes and the halo becomes more symmetric. In Figure~\ref{fig.2}(e), at $t = 5000\, \omega_{pi}^{-1}$, it can be seen that the overall anisotropy of the electron VDF has been considerably reduced and a significant amount of particles is present in the high $v_{\perp}$ - negative $v_{\parallel}$ region of the phase space. 
The combined effect of Landau resonance, late stage of the anomalous cyclotron resonance, and normal cyclotron resonance contributes to the isotropization of the total electron VDF and the formation of a halo population distributed at all pitch angles. These resonant wave-particle interaction mechanisms become predominant during the relaxation phase of the oblique WHFI \citep{Micera2020b}, as the whistler modes shift towards near-parallel propagation angles (see Figures~\ref{fig.4}(e) and (f)). 

We calculated the wave frequency ($\omega_r=0.00622~ \omega_{pi}$), used for the evaluation of the resonant conditions at this time, by solving the linear dispersion relation for the full spectrum of whistler heat flux instabilities (see Section~\ref{seclin-th}). We obtained the parallel wave number ($k_{\parallel}=9~ c/\omega_{pi}$) from the FFT of $\delta B_z$ (see Figure~\ref{fig.4}(e)). We emphasize that as the solar wind propagates away from the Sun it finds more favorable conditions for scattering processes to occur \citep{Lieweretal2001}. Electrons resonate with the whistler waves at gradually higher $n$ and the values of $v_{\parallel}$ at which the anomalous cyclotron resonances occur are becoming closer.

The competition between the perpendicular cooling of the plasma produced by the expansion, and the perpendicular heating and pitch-angle scattering of the strahl carried out by the WHFIs during the first part of the simulation, results in the electron VDF displayed in Figure~\ref{fig.2}(f). At $t = 7000\, \omega_{pi}^{-1}$, the oscillating magnetic energy of the plasma is close to local minimum (see Figure~\ref{fig.1}(a)) and the electron distribution function appears stable with respect to kinetic instabilities. The VDF is characterized by a field-aligned strahl, whose extent, however, is less pronounced than that at the initial time of the simulation, and a suprathermal halo that is present in all directions.
The quasi-stable phase persists until the moment when the second growth of whistler waves, combined with the excitation of the EFI by the core, starts again to reshape the electron VDF. 
In Figures~\ref{fig.2}(g),(h) and (i), where the electron VDF is displayed at $t = 10500$, $14000$ and $16000\, \omega_{pi}^{-1}$, respectively, it is possible to see the reemergence of non-Maxwellian horn-like structures in the high $v_{\perp}$ - positive $v_{\parallel}$ region of the phase space. These features are indicators of ongoing anomalous cyclotron resonant scattering processes.

In Figure~\ref{fig.3} the cuts of the total electron VDF along the parallel ($f_e (v_{\parallel}, v_{\perp} =0)$) and perpendicular ($f_e (v_{\parallel} =0, v_{\perp})$) directions are shown at four different stages of the simulation.
The initial electron VDF in Figure~\ref{fig.3}(a) is characterized by a pronounced field-aligned, anti-sunward directed ``shoulder", which is signature of the presence of a strahl with limited angular extent (at $t=0$, $T_{s \perp}/T_{s \parallel} = 0.5$).
In Figure~\ref{fig.3}(b), at $t = 3000\, \omega_{pi}^{-1}$, the effects of the wave-particle interactions are already visible. The electron VDF starts to deviate from the Maxwellian shape. A portion of the strahl electrons starts to resonate, and then the VDF exhibits bumps corresponding to the ``horns" that represent the seed for the halo formation.
At $t = 5000\, \omega_{pi}^{-1}$ (Figure~\ref{fig.3}(c)), during the non-linear stage of the oblique WHFI, the diffusion of the seed population towards higher values of $v_{\perp}$ has already passed the peak of its activity. The scattered electrons start to relax and are reorganized to form a suprathermal halo distributed at all pitch angles at the expense of the strahl.
The final VDF cuts are depicted in Figure~\ref{fig.3}(d) for $t = 16000\, \omega_{pi}^{-1}$. They show the final state of the VDF after the electron non-thermal features are reshaped by the destabilization and relaxation of the firehose and whistler instabilities in the second part of the simulation as shown in Figure~\ref{fig.2}. 
In comparison with Figure~\ref{fig.3}(a), one can see in Figure~\ref{fig.3}(d) that the number of non-thermal electrons with high $v_{\parallel}$ significantly decreased.   

To analyze the nature of the broad spectrum of the generated waves, in Figure~\ref{fig.4} we display the FFT of the out-of-plane fluctuating magnetic field component $\delta B_z$ at different times.
In Figure~\ref{fig.4}(a), at $t=0$, the $\mathit{k_{\parallel}~ \mhyphen~ k_{\perp}}$ plane shows only the initial numerical noise.
Due to the expansion and hence to the electron kinetic instabilities that are triggered by it, the 2D power spectrum evolves towards the situation shown in Figure~\ref{fig.4}(b), where it peaks at $k_{\parallel} \approx 6 \; \omega_{pi}/c$ and $k_{\perp}$ between $11$ and $14\; \omega_{pi}/c$. This means that at $t = 3000\, \omega_{pi}^{-1}$ the triggered waves present highly oblique angles of propagation with respect to the field, mainly from $61^{\circ}$ to $67^{\circ}$.

When the linear growth phase of the oblique WHFI has already come to the end, and the subsequent relaxation phase of the fastest growing modes starts, the power shifts towards higher $k_{\parallel}$ and smaller propagation angles.
Figures~\ref{fig.4}(c), (d) and (e) show a clear gradual transfer of the power towards the background magnetic field direction. However, the wave magnetic energy at these stages is still distributed at a range of propagation angles rather than being exclusively concentrated along the parallel direction. This implies that during the non-linear phase of the oblique WHFI, resonant scattering of electrons by oblique waves can still occur, in agreement with \citet{Levinson1992} and \citet{Komarov2018}.

At $t= 6000\, \omega_{pi}^{-1}$ (Figure~\ref{fig.4}(f)), the oblique WHFI is approaching the saturated marginal state. The wave power has clearly faded, it is concentrated in a region of the $k_{\parallel} > 8\; \omega_{pi}/c$, and exhibits also a purely field-aligned component that has become comparable to the oblique one. It is during this phase of the simulation that most of the redistribution of the scattered suprathermal electrons takes place to form a symmetric halo.

After the quiet phase that follows the saturation of the WHFIs, the expansion brings the electrons back to unstable conditions. At $t= 10500\, \omega_{pi}^{-1}$ the activity of electromagnetic waves starts again and this time it is initially due to the core crossing the threshold of the resonant EFI (see Figure~\ref{fig.1}(d)). The free energy associated to the electron core temperature anisotropy is converted into the magnetic energy of wave fluctuations. The maximum growth rate of the instability is again mainly in the oblique direction, at an angle about $50^{\circ}$ with respect to $\bm{B}_0$ (Figure~\ref{fig.4}(g)). The identification of these modes as EFI is confirmed by the linear theory, which shows the non-propagating character of this branch (see Section~\ref{seclin-th}).

At $t= 12500\, \omega_{pi}^{-1}$ in Figure~\ref{fig.4}(h) an additional contribution appears in the $\delta B_z$ power spectrum. These modes are also oblique and can be attributed to the excitation of a second surge of WHFIs (see Figures~\ref{fig.1}(b) and (c), and Figures~\ref{fig.4b}(e) and (f) below). Their power is concentrated at higher $k_{\parallel}$ and $k_{\perp}$ than the simultaneous modes produced by the EFI.
Finally, in Figure~\ref{fig.4}(i), where the FFT of $\delta B_z$ is computed at $t= 16000\, \omega_{pi}^{-1}$, we see the damping of both unstable mode waves and their transition towards nearly parallel angles as expected for both oblique EFI \citep{Camporeale2008, Micera2020} and oblique WHFI \citep{Micera2020b}.

\subsection{Kinetic linear theory}\label{seclin-th}

Kinetic linear theory can help to understand the nature of the various wave fluctuations and their impact on the electrons through resonant interactions.
We find numerically the unstable solutions by solving a Vlasov–Maxwell dispersion formalism \citep{Stix1992}, using a root finder based on the Müller’s method.
This has been done by using the dispersion solver DIS-K in the Maxwellian limit \citep{Lopezetal2021} and hence approximating the true electron distribution with a drifting bi-Maxwellian model. The liner dispersion relation confirms that initially our system is not subject to any unstable mode.
We then obtain the general dispersion and stability analysis for the entire wave spectrum at three significant instants of the simulation.

\begin{figure*}
 \centering
 \includegraphics[width=0.75\textwidth]{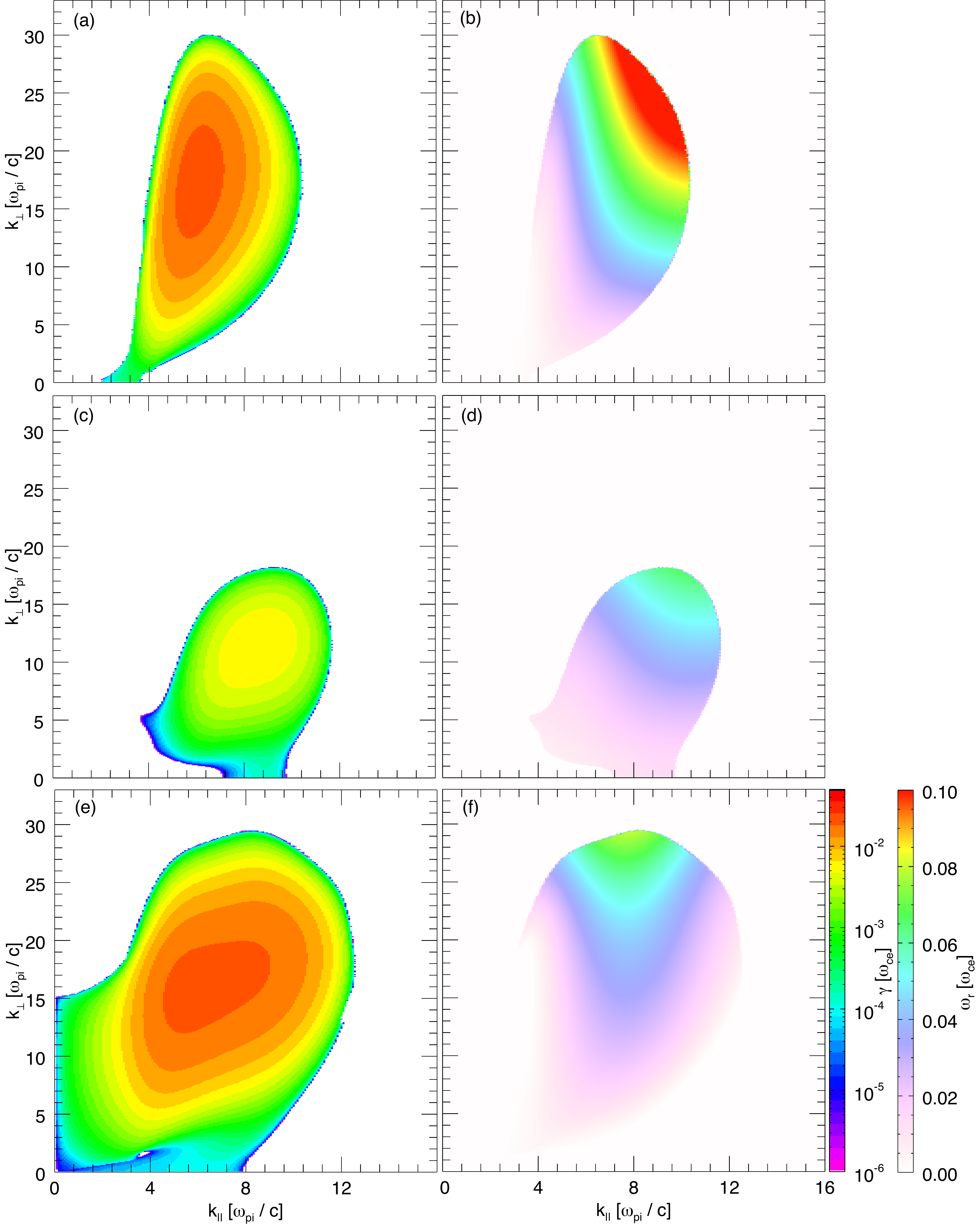}
 \caption{Growth rates $\gamma$ (left column) and corresponding real wave frequencies $\omega_r$ (right column) in the $\mathit{k_{\parallel}~ \mhyphen~ k_{\perp}}$ plane obtained by solving the linear dispersion relation. The results of the calculations are shown at $t = 3000\, \omega_{pi}^{-1}$ for the oblique WHFI linear growth phase (panels (a) and (b)), at $t = 5000\, \omega_{pi}^{-1}$ for the oblique WHFI non-linear relaxation phase (panels (c) and (d)) and at $t = 11250\, \omega_{pi}^{-1}$ for the oblique WHFI + EFI linear growth stage (panels (e) and (f)).
Both growth rates and wave frequencies are in $\omega_{ce}$, with $\omega_{pi} (t = 3000~ \omega_{pi}^{-1}) / \omega_{ce} (t = 3000~ \omega_{pi}^{-1})  = 3.86$, $\omega_{pi} (t = 5000~ \omega_{pi}^{-1}) / \omega_{ce} (t = 5000~ \omega_{pi}^{-1}) = 5.25$ and $\omega_{pi} (t = 11250~ \omega_{pi}^{-1}) / \omega_{ce} (t = 11250~ \omega_{pi}^{-1})  = 7.26$.} \label{fig.4b}
\end{figure*}

In Figures~\ref{fig.4b}(a) and (b) we show the instability growth rate and the real wave frequency provided by the linear theory at $t = 3000\, \omega_{pi}^{-1}$, for the entire range of propagation angles. We confirm that at this time the system is unstable to the right-hand elliptically polarized oblique WHFI \citep{Vasko2019,Verscharen_2019, Lopez2020}: the fastest growing whistler mode is located at $k_{\parallel}\approx6\,\omega_{pi}/c$ and $k_{\perp}\approx16\,\omega_{pi}/c$, has a growth rate $\gamma_\text{max}\approx 0.023\,\omega_{ce} =0.006\,\omega_{pi}$, and a real wave frequency, $\omega_r\approx0.055\,\omega_{ce}= 0.014\,\omega_{pi}$.
We also notice the presence of unstable modes at much lower angles, around $k_{\parallel}\approx3\; \omega_{pi}/c$, $k_{\perp}\approx0$. These modes have a significantly lower growth rate than the oblique whistler fluctuations and correspond to the simultaneous excitation of the left-hand circularly polarized firehose heat flux instability \citep{Shaaban2018MN,Shaaban2018PoP, Lopez2020}. The latter modes exhibit negligible wave frequencies (non-propagating modes with $\omega_r\approx0$).

At $t = 5000\, \omega_{pi}^{-1}$, in Figures~\ref{fig.4b}(c) and (d), the strongest instability for this regime remains the purely oblique whistler heat flux instability, although in its relaxation phase, with maximum growth rate $\gamma_\text{max}\approx 0.005\; \omega_{ce} = 0.001\; \omega_{pi}$ at $k_{\parallel}\approx9\; \omega_{pi}/c$, $k_{\perp}\approx11\; \omega_{pi}/c$, and wave frequency for the fastest growing mode $\omega_r\approx0.033\,\omega_{ce}= 0.006\,\omega_{pi}$.
Nevertheless, propagating whistler modes aligned to the magnetic field (around $k_{\parallel}=9\; \omega_{pi}/c$) are present at this stage, when the oblique WHFI is in its relaxation phase. When these results are compared with the simulation, it should be noted that at $t = 5000\, \omega_{pi}^{-1}$ in Figure~\ref{fig.4}(e) the parallel modes are hardly recognisable, but start to appear a bit later and only become comparable to the oblique modes at $t = 6000\, \omega_{pi}^{-1}$ (see Figure~\ref{fig.4}(f)).

When we solve the linear dispersion relation for the plasma parameters at $t = 11250\, \omega_{pi}^{-1}$ we observe a combination of oblique left-hand non-propagating modes with $\omega_r \approx0$, typical of the resonant electron firehose instability \citep{LiandHabbal2000, Camporeale2008}, and 
once again propagating whistler modes with $\omega_r \ne 0$ (Figures~\ref{fig.4b}(e) and (f)). The EFI, which starts as soon as the core electrons approach the instability threshold, as shown in Figure~\ref{fig.1}(d), excites modes characterized by wave numbers that are slightly lower than those generated at the same time by the oblique WHFI. As a result, the region of the wave spectrum characterized by low $k_{\parallel}$ and $k_{\perp}$ also has the frequency values of $\omega_r \approx0$ (bottom-left corner of Figures~\ref{fig.4b}(f)). The modes with $k_{\parallel}> 6\, \omega_{pi}/c$ present the frequency range typical of the whistler waves that we have simulated, which is also the whistler frequency range commonly observed in the near-Sun solar wind (below $0.1\; \omega_{ce}$) \citep[e.g.][]{Agapitov2020, Jagarlamudi2021}.

The results obtained from the kinetic linear theory are in good quantitative agreement with those inferred from the simulation. The linear theory clearly captures the oblique whistler modes responsible for the substantial deformation of the electron VDF, identifies their transition to reduced propagation angles, and finally, manages to distinguish the quasi-overlapping whistler and firehose modes that are generated in the last phases of the simulation. However, there are some minor differences between the linear theory results and the FFT obtained from our simulation. They are essentially due to the use of the Maxwellian distribution functions to describe the electrons in the theory, while in the simulations they show clear departures from this model.

\subsection{Solar wind heat flux}
The evolution of the heat flux in the heliosphere can provide information about how the solar wind is loaded with photospheric energy and how this energy is redistributed as the wind propagates away from the Sun.
The heat flux is strongly regulated by kinetic phenomena and in particular by wave-particle interaction processes \citep{Bale_2013, Halekas2020, Cattell2021}. As the solar wind expands, it encounters different plasma conditions and hence regimes characterized by different heat flux regulation processes.

The solar wind heat flux is almost exclusively carried by electrons, and in particular by the suprathermal electron populations \citep{Feldman1975, Pilipp1987}. In the inner heliosphere, where the suprathermal portion of the electron VDF is clearly dominated by the strahl \citep{Halekas2020, Bercic2020}, it is this population of electrons that is the main responsible for the heat flux in the solar wind. Given its marked anisotropy, i.e. its being flattened along the direction of the magnetic field, the strahl carries a heat flux with predominantly radial component. 
For these reasons, the heat flux carried in the reference frame of the solar wind corresponds to that carried by the strahl along the direction parallel to $\bm{B_0}$.

The heat flux can be defined as $Q_{s \parallel} = \frac{m_s}{2} \int v_{\parallel} v^2 f_s d^3v$, and split in three components \citep{Feldman1975, Scime1994}:
$Q_{s} = Q_{\text{enth}, s} + Q_{\text{bulk}, s} + q_s$,  with $Q_{\text{enth}, s} = \frac{3}{2}~ n_s m_s u_s w_s^2 $ the strahl electron enthalpy, $Q_{\text{bulk}, s} = \frac{1}{2}~ m_s n_s u_s^3$ the energy flux associated to the bulk motion of the electrons, and $q_{s} = \frac{m_s}{2} \int (v_{\parallel} - u_s) (v - u_s)^2 f_s d^3v$ the heat flux carried by the strahl in its frame of reference (skewness of the VDF).

\begin{figure*}
 \centering
 \includegraphics[width=1\textwidth]{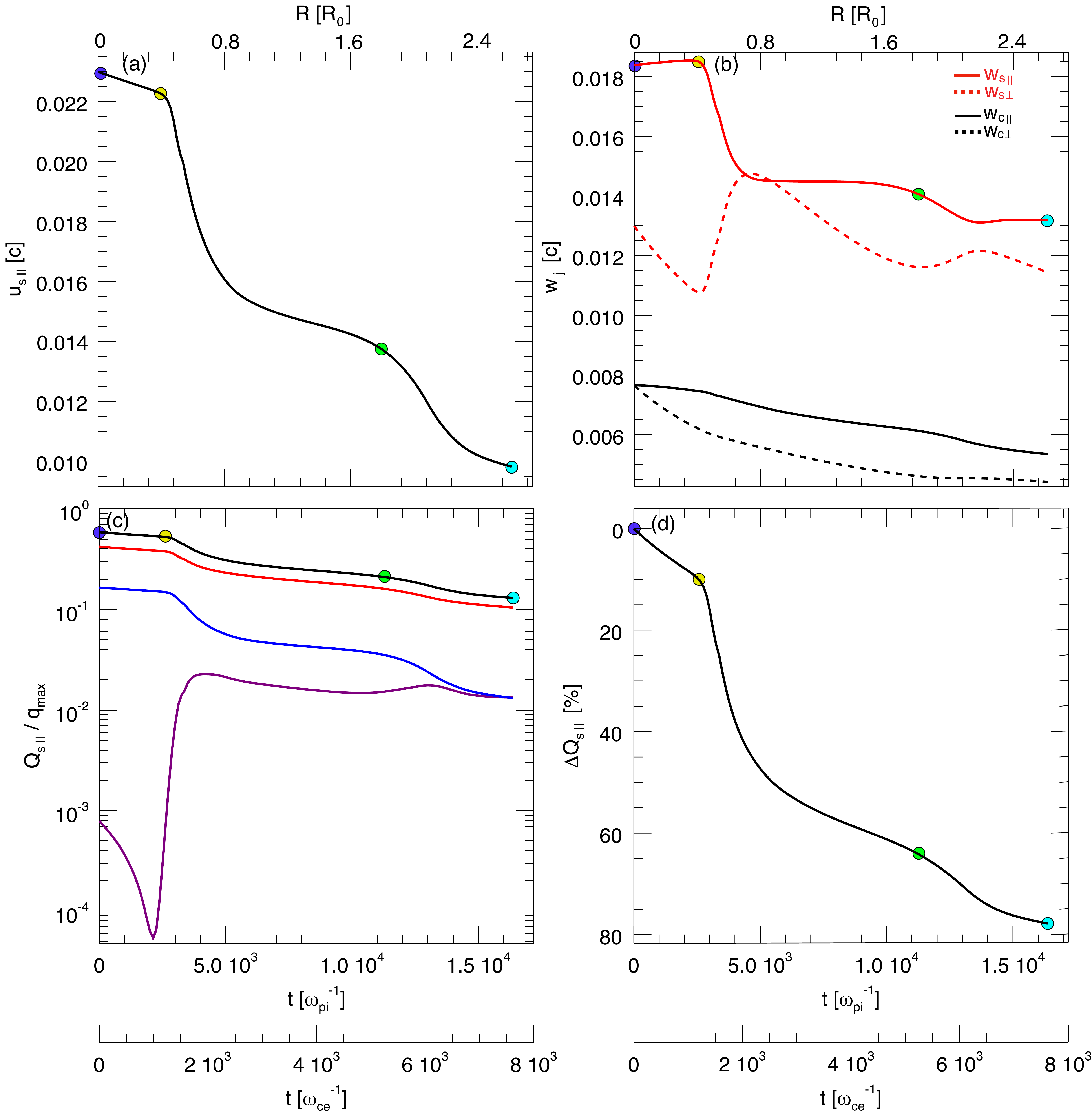}
 \caption{Temporal evolution of the parallel strahl drift velocity normalized to the speed of the light in vacuum (panel (a)).
 Parallel (solid lines) and perpendicular (dashed lines) thermal velocities of the strahl (red) and core (black) electrons as a function of time (panel (b)).
 Evolution of the heat flux components carried by the strahl along the magnetic field direction: $Q_s$ (black), $Q_{\text{enth}, s}$ (red), $Q_{\text{bulk}, s}$ (blue), and $q_{s}$ (purple). All the heat flux components are normalized to $q_{max} = \frac{5}{2}~ m_e (n_c w_{c}^3 + n_s w_{s} ^3)$ (panel (c)).
Percentage decrease of the total electron heat flux as a function of time ((panel (d)).
In all the four panels, the upper axis shows the heliocentric distance $R/R_0$, while the two bottom axes show the time in units of $\omega_{pi}^{-1}$ and $\omega_{ce}^{-1}$. The initial time is marked with a blue dot, the first oblique WHFI onset ($t = 2500\, \omega_{pi}^{-1}$) with a yellow dot, the oblique WHFI + EFI second growth stage ($t = 11250\, \omega_{pi}^{-1}$) with a green dot, and the final simulation time with a cyan dot, similarly to Figure~\ref{fig.1}.} \label{fig.5}
\end{figure*}

In agreement with \citet{Innocenti2020} and \citet{Micera2020b}, one of the main factors that leads to the reduction of the heat flux is the rapid decrease of $u_{s \parallel}$, which is in turn due to the combined effect of the interconnected expansion and development of electron-scale instabilities.
In Figure~\ref{fig.5}(a) we show the temporal evolution of the parallel strahl drift velocity normalized to the speed of the light in vacuum, $u_{s \parallel}/c$ (cf. Figure~\ref{fig.1}(b), where the strahl drift velocity is normalized to the local Alfvén velocity). It can be seen that, after the first phase of undisturbed expansion, $u_{s \parallel}/c$ has a sudden decrease in correspondence with the onset of the oblique WHFI. When this first oblique WHFI is saturated, the parallel strahl drift velocity decreases less rapidly, but this only lasts until the time when the core EFI and the second stage of the oblique WHFI are triggered and further considerably decrease its value.

The evolution of the parallel and perpendicular components of the core and strahl thermal velocities can be seen in Figure~\ref{fig.5}(b), which shows the trends due to the successive phases of excitation and relaxation of the electron kinetic instabilities.
The pitch-angle scattering of the electrons during the development of the WHFIs leads to a substantial redistribution of the strahl thermal energy from the parallel to the perpendicular direction and hence to a broadening of the strahl pitch-angle distribution. The process acts as a driving force for the halo formation. Due to the generation of whistler waves, the perpendicular cooling produced by the expansion has a limited impact not only on the strahl, but also, to a lesser extent, on the core. Already at the first onset of the oblique WHFI, the core shows a change of slope in the evolution of both parallel and perpendicular components of $w_{c}$. 
A further reduction of the core temperature anisotropy occurs as expected in correspondence with the EFI growth phase. 

Both the kinetic energy redistribution and the drop of the strahl drift velocity are reflected in the evolution of the electron heat flux. In Figure~\ref{fig.5}(c) we show the field-aligned heat flux associated with the strahl in the rest frame of the solar wind and the three terms in which it can be split, as functions of time. All the heat flux components are normalized to the saturation heat flux $q_{max} = \frac{5}{2}~ m_e (n_c w_{c}^3 + n_s w_{s} ^3)$. 
Both enthalpy and bulk components of the heat flux exhibit a sudden and steep decrease as soon as whistler waves start to be generated. The total heat flux presents the characteristic multi-slope behavior seen also in other electron bulk parameters.
Figure~\ref{fig.5}(c) also shows that the heat flux is mainly carried by the suprathermal electron enthalpy, but the component that is affected by the collisionless regulation more significantly is the one related to the bulk motion, in agreement with \citet{Innocenti2020} and \citet{Micera2020b}.
The electron heat flux in the rest frame of the strahl, $q_s$, presents first a rapid growth simultaneous with the oblique WHFI onset, when the VDF is more asymmetric, and a second growth, more moderate, at the EFI onset.

In Figure~\ref{fig.5}(d) we quantify the heat flux reduction, by showing the evolution of its percentage decrease. At the end of our simulation, the solar wind has dissipated about $80\%$ of the heat flux that it was carrying at the initial time step. Almost a half of it was dissipated during the first interaction stage between the electron VDF and the whistler waves. The heat flux percentage decrease shows the typical multi-trend behaviour corresponding to the effectiveness of the repeated wave-particle interactions.

\section{Discussion}\label{sec.4}
\subsection{Comparison with observations}
We now compare our results to recent observations. 
The implications of our simulation can help to clarify the dynamics of electrons in the solar wind.
We have confirmed the effects of the whistler heat flux instabilities on the electron VDF \citep{Micera2020b}, namely the scattering of the strahl electrons into the halo \citep{Stevark2008} and the significant heat flux suppression simultaneous to the instability development \citep{Halekas2020b}. 

We have demonstrated how the expansion can act as a driver for the excitation of whistler instabilities and hence as a self-sustaining mechanism for the generation of whistler waves, which seem to have many aspects in common with those detected in the near-Sun solar wind by PSP.
The fastest growing whistler-mode waves triggered in our simulation present angles of propagation with respect to the background magnetic field from $61^{\circ}$ to $67^{\circ}$ (Figure~\ref{fig.4}(b)). This is in qualitative agreement with the highly oblique whistler waves detected during the PSP Encounter 1 and reported by \citet{Agapitov2020} and \citet{Cattell_2021_ArXive}.

The magnetic amplitude of such highly oblique whistler waves reaches values up to $\lvert \delta B \rvert  \sim 0.5~ B$ (Figure~\ref{fig.1}(a)). 
These values are consistent with large-amplitude whistler waves reported by \citet{Agapitov2020} and \citet{Cattell_2021_ArXive} in the near-Sun solar wind and with those observed in the STEREO electric field data \citep{Breneman_2010, Cattell2020}, but are significantly higher than the amplitudes of parallel whistler waves observed in Cluster and THEMIS search coil data, which are typically less than $0.02~ B$ \citep{Lacombe2014, Tong2019}.
In addition, in line with the results by \citet{Cattell_2021_ArXive}, the peak amplitude of the simulated whistlers does not change much with the heliocentric distance (see Figure~\ref{fig.1}(a)). This suggests that the whistlers' effect on the solar wind electron VDF may occur over a wide region of the heliosphere.

The values of the whistler wave frequency in our simulation fall in the observed range as reported by \citet{Agapitov2020}, \citet{Cattell_2021_ArXive}, \citet{Jagarlamudi2021}. Indeed, according to these works, the whistler perturbations in the inner heliosphere have frequencies ranging from about $0.03$ up to $0.2~\omega_{ce}$ (cf. Figure~\ref{fig.4b} above).

The perpendicular electron heating driven by the whistler waves limits the development of the temperature anisotropy produced by the expansion \citep{Innocentietal2019}. The temperature of the suprathermal electron populations, as well as that of the core, diverge from the double adiabatic expansion law \citep{Chew1956} already when the first oblique WHFI is triggered and way before reaching the EFI threshold (see Figures~\ref{fig.1}(c) and (d) and Figure~\ref{fig.5}(b)). This suggests that whistler heat flux instabilities are a key mechanism to explain the limited values of electron temperature anisotropy observed in the solar wind, on a par with the firehose and whistler temperature anisotropy instabilities that are most commonly considered in this respect.
Similarly to observations of the halo reported by \citet{Stevark2008}, our simulation shows that the suprathermal portion of the electron VDF (strahl and halo), despite the expansion, never really approaches the threshold of the firehose instability. It is rather concentrated in the stable region at the center of the electron $\beta_{\parallel}$ vs $T_{\perp}/T_{\parallel}$ plane (so-called ``Brazil plot", see Figure~\ref{fig.1}(c)). This plausibly means that, when looking exclusively at the classical ``Brazil" plot for the core and for the suprathermal electrons, one does not have the full information on which instability limits the temperature anisotropies. This is particular relevant for the suprathermal component of the electron VDF \citep{Stevark2008, Bercic2019}.

The whistler waves also produce perpendicular heating of the core, which causes the core temperature anisotropy evolution to deviate from that expected from a double adiabatic expansion.
This corresponds to observations reported by \citet{Jagarlamudi2020} and \citet{Cattell2021}, according to which during the intervals of whistler wave detection there seems to be a concomitant increase of the core electron $\beta$.

\subsection{Origin of field-aligned sunward whistler waves}\label{sect_4.2}
In this Section, we propose a mechanism for the self-generation of whistler waves aligned with the background magnetic field, propagating mainly in the sunward direction. They may be the result of the non-linear evolution of the oblique WHFI and of the shift of the oblique whistler waves towards small propagation angles.

\begin{figure}[] 
\centering
\includegraphics[scale=0.43]{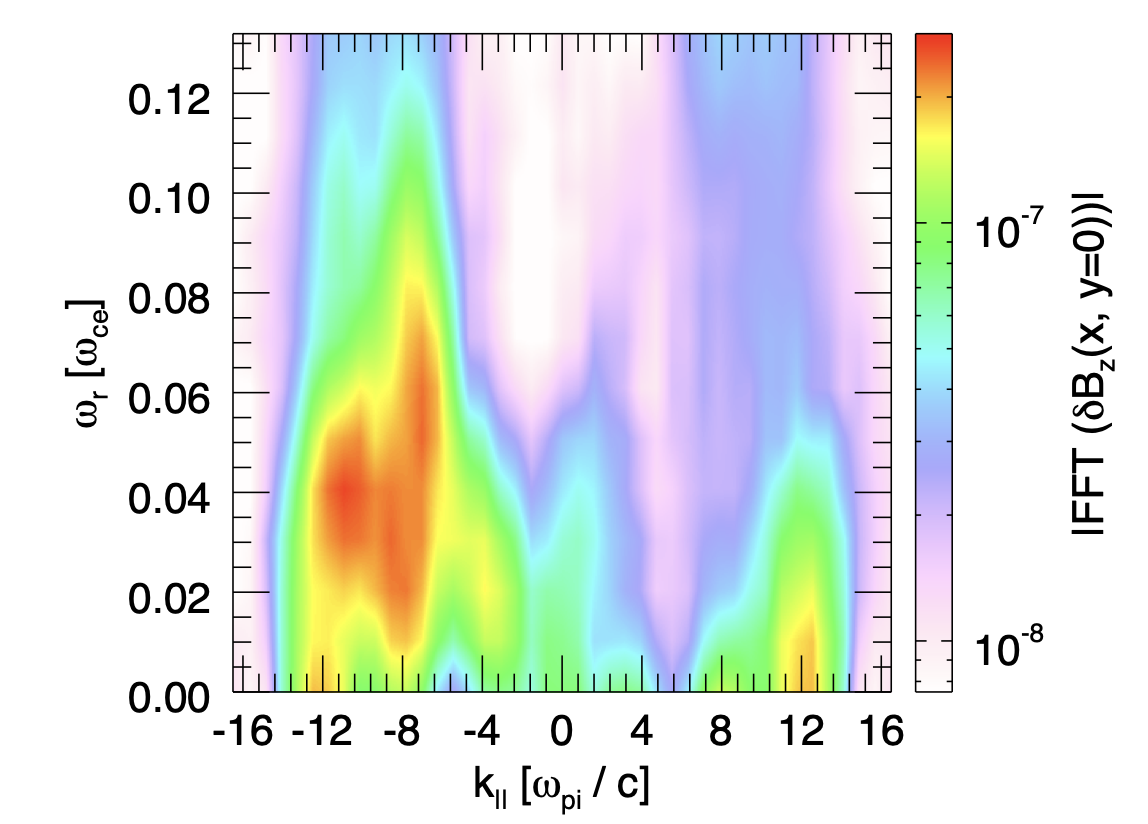}
\caption{Space-time Fourier power spectrum ($k_{\parallel} - \omega_r$) of both sunward ($k_{\parallel} < 0$) and anti-sunward ($k_{\parallel} > 0$) whistler waves propagating along the background magnetic field direction for the interval $5000 < t\, \omega_{pi}  < 8000$.}\label{fig.7}
\end{figure}

In Figure~\ref{fig.7} we show the space-time Fourier power spectrum, in the $k_{\parallel} - \omega_r$ plane, obtained exclusively for field-aligned whistler-mode waves. It was computed as the FFT of the cut along the parallel direction of the out-of-plane component of the fluctuating magnetic field, $\delta B_z (x, y =0)$. We have selected a time interval ranging from $t = 5000\, \omega_{pi}^{-1}$ to $t = 8000\, \omega_{pi}^{-1}$ (relaxation phase of the oblique WHFI). As shown in Figure~\ref{fig.4}, this is the stage of the simulation in which the unstable modes are no longer concentrated only at high propagation angles, but instead the field-aligned component has become significant. We see that during the selected interval there is a considerable portion of whistler waves propagating in the sunward direction and hence anti-parallel to the electron heat flux. The fastest growing parallel and anti-parallel whistler waves are characterized by wave numbers between $8$ and $12\; \omega_{pi}/c$, confirming those shown in Figure~\ref{fig.4}(f), and have frequencies below $0.1\; \omega_{ce}$, in line with those depicted in Figure~\ref{fig.4b} and with in situ observations \citep[e.g.][]{Agapitov2020, Cattell_2021_ArXive, Jagarlamudi2021}.

This result has a fundamental importance. Indeed, for a long time it was considered that most of the whistler waves in the solar wind, and consequently most of the collisionless heat flux regulation, were due to the parallel WHFI, which has its maximum growth rate in the direction of the magnetic field \citep{Gary1975, Gary1994}. In support of this, \citet{Tong2019b} made a statistical analysis of the occurrence of whistler waves around 1 au and showed that there is a very good similarity between the properties of the observed waves and those expected from the waves generated by the parallel WHFI.

Nevertheless, it remains unexplained how the whistler waves generated by the parallel WHFI instability can regulate the electron heat flux. Indeed, it has been demonstrated through both linear theory \citep{Shaaban2019d,Vasko2020} and simulations \citep{Kuzichev2019, Lopez2019} that these waves travel only parallel to the heat flux and consequently are unable to scatter the suprathermal component of the electron VDF.

Our simulation suggests that the relaxation of the oblique WHFI is a plausible source of whistler waves propagating at limited angles with respect to the magnetic field and, more importantly, mainly in the sunward direction. These anti-parallel waves are able to interact with the filed-aligned strahl and thus to explain the significant non-collisional contribution to the heat flux regulation  \citep{Salem2003, Bale_2013, Halekas2020b}.

We suggest that the fast (compared to the expansion time scales) and continuous cycle of excitation and relaxation of the oblique WHFI in the inner heliosphere produces the whistler waves with propagation angles ranging from parallel to oblique, and the portion of parallel whistlers propagating sunward, in agreement with observations \citep{Agapitov2020, Cattell_2021_ArXive}. 
This mechanism of whistler wave generation by the expansion can essentially occur in the inner heliosphere. As the distance from the Sun increases, on the one hand, the relative velocity between the core and the suprathermal component of the electron VDF decreases due to the combined effect of kinetic instabilities and solar wind expansion \citep[e.g.][]{Innocenti2020}. On the other hand, $v_A$ stops to decrease steeply when the transverse component of the magnetic field becomes dominant \citep{Chew1956}: when the field is radial, then $v_A \sim R^{-2}/R^{-1} \sim R^{-1}$; when the azimuthal component dominates, then $v_A \sim R^{-1}/R^{-1} = const$.
The combination of these two factors significantly inhibits the development of the oblique WHFI \citep{Lopez2020}, which requires values of $u_s /v_A$ above a certain threshold to be triggered.

At larger heliocentric distances, the question seems to be more intricate. Regarding the parallel small-amplitude and commonly anti-sunward whistler waves \citep{Lacombe2014, Stansby2016, Tong2019}, they can be explained by the fact that with the increase of the heliocentric distance, the suprathermal component of the electron VDF becomes progressively more isotropic \citep[e.g.][]{Bercic2019} so that the plasma satisfies the conditions for the occurrence of the parallel WHFI instability \citep[see][]{Kuzichev2019, Lopez2019, Micera2020b}. As far as the highly oblique and large-amplitude whistler waves observed by \citep{Breneman_2010} and \citep{Cattell2020} are concerned, the question of their origin remains open. Around 1 au, where the expansion does not increase the $u_s / v_A$ ratio anymore, there may be other phenomena that can significantly modify the structure of the electron VDF and bring the plasma into unstable conditions. According to \citet{Cattell_2021_ArXive}, at 1 au oblique whistler waves are often associated with stream interaction regions or even coronal mass ejections that may give rise to conditions similar to those reproduced by \citet{Micera2020b}, where the instability is independent of the presence of the azimuthal magnetic field component, and the large value of $u_s$ is the main driver of the instability.

It is important to emphasise a crucial difference between the presented results and those obtained by \citet{Micera2020b} in a non-expanding case study. The process of relaxation of oblique modes towards parallel angles can be seen in both expanding and non-expanding simulations. However, in the simulation reported by \citet{Micera2020b}, the electron VDF, and in particular its suprathermal component, became almost isotropic after the scattering on the oblique whistler waves and remained almost unperturbed until a new instability was triggered: the resulting VDF was indeed suitable for the excitation of the parallel WHFI. In addition to the relaxation of oblique modes towards parallel angles, long after the relaxation of the oblique WHFI, new modes appeared, which were exclusively anti-sunward directed and unrelated to the oblique WHFI modes. 
In the present work, the expansion of the solar wind has the effect of continuously bringing the electrons into anisotropic conditions. The presence of anisotropic suprathermal electron populations inhibits the growth of the parallel WHFI, and the high $k_{\parallel}$, exclusively anti-sunward directed modes, which are typical of parallel WHFI, are not seen in the expanding simulation. Therefore, the only modes with a significant growth and propagating along the magnetic field direction are those resulting from the relaxation of the oblique WHFI, which manifest also a significant sunward component.
Our results confirm the need to take into account the solar wind expansion in order to have a more complete picture of the evolution of the electron VDF during its interaction with whistler waves.

\subsection{Limitations of our approach}
We now comment on the approximations adopted in our numerical model.
In order to reduce the computational cost of our simulation, we significantly decreased the expansion time scale with respect to realistic ones. 
If one considers a heliocentric distance $R=30~ R_\odot$ and the wind speed  $u=600$ km/s, the resulting realistic expansion time will be $\tau_r = R/u \sim 10^4~ $s. If this value is compared with our $\tau_{exp} = 0.4$ s, then $\tau_{exp} / \tau_r \sim 10^{-5}$. This means that the expansion in our simulation has been sped up by a factor $10^{5}$ with respect to the realistic expansion times.
However, this is acceptable as the expansion time scale is longer than the time scales of triggered instabilities. With the selected expansion time we have $\tau_{exp} / \tau_{~WHFI}\approx 10$ and $\tau_{exp}/ \tau_{~OWHFI}\approx 60$, with $\tau_{~WHFI}$ and $\tau_{~OWHFI}$ evaluated as the inverse of the maximum growth rates of the simulated parallel and oblique WHFIs, respectively (see Section \ref{seclin-th}). This confirms that the selected $\tau_{exp}$ value is within the limits of validity of the expanding box model and gives us confidence about the physical significance of our results.

The speedup of the expansion effects leads to the electron core firehose instability triggered almost simultaneously with the second cycle of destabilization of the oblique WHFI. 
The compression of the temporal scales that separate the occurrence of these two instabilities intensifies their effects on the electron VDF and hence on the heat flux that the electrons carry. 
The heat flux carried by the electrons is almost entirely dissipated while the wind travels a heliocentric distance of only $R=2.65~R_0$, which is clearly not the case in reality. 

We also note that the background magnetic field in our simulation is purely radial. This causes the Alfvén velocity to decrease as $v_A \sim R^{-1}$ throughout the entire simulated time frame (as $B_r \sim R^{-2}$ and $n \sim R^{-2}$), causing in turn the oblique WHFI to be destabilized repeatedly during the solar wind expansion. However, this behavior is only expected in the inner heliosphere. As the solar wind moves away from the Sun, the transverse component of the magnetic field $B_t \sim R^{-1}$ becomes dominant and causes the Alfvén velocity to become almost constant. This means that the ratio $u_s/v_A$ does not increase anymore and, together with the depletion of the strahl caused by the oblique WHFI itself, implies that this instability is no longer excited \citep{Verscharen_2019, Lopez2020}, and the simulated cycles of stabilization and destabilization come to an end.

Finally, we argue that in the expansion phase that precedes the development of the microinstabilities, our plasma follows a double adiabatic evolution, which, however, does not appropriately describe the true evolution of the solar wind where collisionless instabilities, turbulence and collisions are ubiquitous \citep{Landi2012}. An approach involving all these processes together goes beyond the scope of this work and will be investigated in future simulations.

\section{Conclusions}\label{sec.5} 

In this work, we used an electron VDF which faithfully reproduces those recently observed during the PSP Encounter 1 \citep{Halekas2020,Bercic2020}, and we followed its evolution with heliocentric distance, in order to shed light on the kinetic processes that influence the electron dynamics during solar wind propagation from the inner heliosphere onward.
This was carried out by performing a two spatial dimensions, three velocity components (2D3V) fully kinetic Expanding Box Model simulation through which we could monitor the nonlinear development of kinetic instabilities and study their effects on the solar wind electrons. The results of the simulations have been compared with those obtained through kinetic linear theory and discussed within the context of recent PSP observational studies \citep[e.g.][]{Agapitov2020, Cattell_2021_ArXive, Cattell2021, Jagarlamudi2021}.

It has been demonstrated that the expansion of the solar wind can be considered as one of the main drivers of the oblique whistler heat flux instability. This instability leads to the generation of whistler waves with main properties being in qualitative agreement with those observed in the inner heliosphere by PSP. It has also been confirmed that these waves, which begin to propagate predominantly in the oblique direction with respect to the ambient magnetic field, can significantly modify the shape of the electron VDF, leading to the formation of the halo at the expense of the strahl, and consequently regulate the electron heat flux.
In addition, due to the continuing expansion, the plasma crosses the threshold of electron kinetic instabilities multiple times, triggering a cycle of excitation and saturation of whistler heat flux and firehose instabilities. This leads to the continuous destabilization of waves which interact with the electron VDF during the entire simulation.

PiC simulations have recently shown that the whistler waves produced by whistler heat flux instabilities can strongly scatter the electron suprathermal populations \citep{Roberg-Clark2018, Komarov2018, Micera2020b}. One of the crucial differences of our study with respect to these earlier works is that our initial electron VDF is completely stable to microinstabilities and naturally evolves towards unstable conditions due to solar wind expansion.

As already shown by \citet{Innocenti2020}, solar wind expansion can affect the heat flux regulation via a two-step process: by modifying the evolution of collisionless instabilities that in turn affect the heat flux regulation. In the present work, we verify this assumption using a more appropriate description for the strahl and 2D simulation geometry, which allow for oblique whistler heat flux and firehose instabilities to develop. We demonstrated that the evolution of heat flux regulating instabilities is modified by the expansion (cf. the results reported by \citet{Micera2020b}) and that this has consequences for the heat flux evolution.

We have also shown that sunward-directed parallel whistler waves can result from the relaxation of oblique whistler waves. This can help to solve the apparent controversy that the field-aligned whistler waves generated by WHFI propagate only parallel to the heat flux direction \citep{Kuzichev2019, Lopez2019, Vasko2020} and fail to interact with the strahl, despite non-collisional phenomena shaping the non-thermal features of the electron VDF being ubiquitous in the solar wind \citep[e.g.][]{Salem2003, Halekas2020b}.

Our work represents an important step in clarifying the role of small-scale electron kinetic processes in the broader context of solar wind physics. It adds new insight into viable mechanisms for heat flux regulation by collisionless processes, by including the effects of plasma expansion in fully kinetic simulations of relevant instabilities. Our results are consistent with several key trends observed in the heliosphere and may provide an explanation for the variation of the average whistler wave propagation angle with heliocentric distance. The parameters of the simulated whistler waves, such as frequencies and amplitudes, are in remarkable agreement with observational data.
The further approach of PSP to the Sun and its observational campaigns coordinated with those carried out by Solar Orbiter during the radial alignment of the two spacecraft, will be able to provide us with information on even more pristine electron VDFs and their radial evolution. This will bring fundamental insight regarding the complementary evolution of kinetic instabilities and electron energy fluxes.

\acknowledgments
A.M. acknowledge L. Matteini for useful discussions and L. Franci and E. Papini for helpful suggestions on theoretical and numerical aspects.
This work was supported by a PhD grant awarded by the Royal Observatory of Belgium to one of the authors (A. M.).
These simulations were performed on the supercomputers SuperMUC (LRZ) and Marconi (CINECA) under PRACE allocations.
A. N. Z. thanks the European Space Agengy (ESA) and the Belgian Federal Science Policy Office (BELSPO) for their support in the framework of the PRODEX Programme. 
R.A.L acknowledges the support of ANID Chile through FONDECyT grant No. 11201048.
This research was supported in part by the NASA DRIVE HERMES project, grant No. 80NSSC20K0604.

\bibliographystyle{aasjournal}  
\bibliography{main}

\end{document}